\documentclass[conference]{IEEEtran}

\usepackage{cite}
\usepackage{times}
\usepackage{subfigure}
\usepackage{nicefrac}
\usepackage{graphicx}
\usepackage{verbatim}
\usepackage{amsmath}
\usepackage{array}
\usepackage{color}
\usepackage{url}
\usepackage{float}
\usepackage{multirow}
\usepackage{caption}

\usepackage{algorithmic}

\begin{document}

\title{
Energy-aware Dynamic Resource Allocation in Virtual Sensor Networks}

\author{\IEEEauthorblockN{Carmen Delgado\IEEEauthorrefmark{1},
Mar\'ia Canales\IEEEauthorrefmark{1},
Jorge Ort\'in\IEEEauthorrefmark{1}\IEEEauthorrefmark{2},
Jos\'e Ram\'on G\'allego\IEEEauthorrefmark{1},\\
Alessandro Redondi\IEEEauthorrefmark{3},
Sonda Bousnina\IEEEauthorrefmark{3},
Matteo Cesana\IEEEauthorrefmark{3}}

   \IEEEauthorblockA{\small \IEEEauthorrefmark{1}Arag\'on Institute of Engineering Research.  Universidad de Zaragoza, Zaragoza, Spain.   \{cdelga, jrgalleg, mcanales\}@unizar.es} 
    \IEEEauthorblockA{\small \IEEEauthorrefmark{2}Centro Universitario de la Defensa Zaragoza. Academia General Militar, Zaragoza, Spain. jortin@unizar.es}
 \IEEEauthorblockA{\small \IEEEauthorrefmark{3} Dipartimento di Elettronica, Informazione e Bioingegneria, Politecnico di Milano, Milano, Italy.\\ \{alessandroenrico.redondi, sonda.bousnina, matteo.cesana\}@polimi.it}    

\vspace{-7mm}			       

  }
\maketitle

\begin{abstract}
Sensor network virtualization enables the possibility of sharing common physical resources to multiple stakeholder applications. This paper focuses on addressing the dynamic adaptation of already assigned virtual sensor network resources to respond to time varying application demands. We propose an optimization framework that dynamically allocate applications into sensor nodes while accounting for the characteristics and limitations of the wireless sensor environment. It takes also into account the additional energy consumption related to activating new nodes and/or moving already active applications. Different objective functions related to the available energy in the nodes are analyzed. The proposed framework is evaluated by simulation considering realistic parameters 
from actual sensor nodes and deployed applications to assess the efficiency of the proposals.
\end{abstract}

\section{Introduction}

The Internet of Things (IoT) paradigm considers that real world objects can be equipped with sensing capabilities to gather information on their environment and then to deliver it to the Internet. This data delivery can be done through wireless multi-hop paths leveraging the cooperation of other smart objects for traffic relaying and making up networks, which are often known as Wireless Sensor Networks (WSN). Typically, the hardware and network resources in WSNs are designed and deployed to the specific application requirements. While this paradigm allows to have ``optimal'' performance on the specific application, it prevents other applications from reusing the already deployed hardware and software resources, leading to the proliferation of redundant WSNs deployments.

In this context, virtualization is a promising technique to achieve an efficient reuse of general purpose wireless sensor networks to dynamically support multiple applications and services  \cite{Khan15}. The key idea behind this approach, which often goes under the names of Virtual Sensor Networks (VSN), is to abstract away ``physical  resources'' including node processing/storage capabilities, available communication bandwidth and routing protocols, which can then be ``composed'' at a logical level to support usage by multiple independent users and even by multiple concurrent applications. This new paradigm has stimulated research efforts in the field of novel programming abstractions at the node level and management framework at the network level to support multiple applications over a shared physical infrastructure \cite{Khan15}, \cite{Madria14}.

Nevertheless, comprehensive solutions for dynamic resource allocation that cope with the specific limitations of WSNs still need to be found. In \cite{Xu2010} an optimization framework for environmental monitoring applications is proposed that aims to perform an application-to-sensors assignment which minimizes the variance of the sensed data. The authors of \cite{Zeng2015TC} propose an optimization framework to prolong network lifetime by properly scheduling the tasks in a shared/virtual sensor network. In previous works \cite{Delgado2015},\cite{Delgado2016} we have proposed a mathematical programming framework to optimally allocate in a static scenario the shared physical resources of the general purpose WSN to multiple concurrent applications. In these works the whole set of applications was known in advance and constant during the analysis time period. Now we focus on addressing the dynamic adaptation of the allocated resources when applications demands are time varying. To respond to increasing network demands, new resources may be allocated or already used resources may have to be reallocated. This implies additional energy consumption related to activating new nodes and/or moving already active applications that must be considered. Due to the energy limitations of sensor networks, we tackle the problem of dynamic resource allocation trying to optimize the use of network energy.  
Numerical results are then obtained by applying the proposed framework to realistic WSN instances to assess the efficiency of the different proposals.

\section{System model and optimization framework}
\label{model}

Let $S = \left\{s_1,s_2, \dotsc , s_l\right\}$ be a set of sensor nodes scattered in a reference area. Each sensor node $s_i$ has an energy budget $E_i$. Let $T = \left\{t_1, \dotsc ,t_n \right\}$ be a set of test points in the reference area, which are physical locations where some parameters must be measured. Let $A = \left\{a_1,a_2, \dotsc , a_m\right\}$ be a set of applications to be deployed in the system. Each application $a_j$ arrives at time $\tau_j$ and has a lifetime $\epsilon_j$. 
In the following, we will use the subscript index $i$ (or $h$) to refer to a sensor node $s_i$ (or $s_h$), the subscript index $j$ to refer to an application $a_j$ and the subscript index $k$ to refer to a test point $t_k$. 

Each application $j$ requires to sense a given set of test points $T_j \subseteq T$. Formally, the application $j$ has to be deployed in a subset of the sensor node set $S$, such that all the test points in $T_j$ are sensed. A test point is covered by a sensor node $i$ if it is within its sensing range, $R_i^{s}$.  Thus, given a test point, a set of sensor nodes can \textit{cover} it, but only one sensor node will \textit{sense} it. Let $S_{jk}$ be the set of sensor nodes which cover the test point $k$, with $k \in T_j$. A necessary condition for an application $j$ to be successfully deployed is that all the test points in its target set $T_j$ must be sensed during its whole lifetime $\epsilon_j$. Each application $j$ in $A$ is further characterized by a characteristic vector $r_j = \left\{c_j, m_j, l_j \right\}$ which specifies the required source rate, memory and processing load demanded by the application when it is deployed on a sensor node. Each sensor node $i$ in $S$ is characterized by a resource vector $o_i = \left\{C_i, M_i, L_i, E_i\right\}$, which specifies its available bandwidth, storage capabilities, processing power and energy.
A protocol interference model with power control is used to characterize the wireless communications. The maximum transmission power is $P_{max}$. 
Given a directional link between a pair of nodes $\left(i,h\right)$, the channel gain from $i$ to $h$ is $g_{ih} = g_0 \cdot d_{ih}^{-\gamma}$, being $d_{ih}$ the distance, $\gamma$ the path loss index and $g_0$ a constant dependent on antenna parameters. If $p_i$ is the transmission power assigned to node $i$, a transmission towards $h$ is successful if $p_i \cdot g_{ih} > \alpha$ and interference at other node is non-negligible if $p_i \cdot g_{ih} > \beta$, being $\alpha$ and $\beta$ the receiver and interference sensitivities. Thus, the transmission range for node $i$ can be obtained as $R_i^T\left(p_i\right)=\left(p_i \cdot g_0/\alpha\right)^{1/\gamma}$ and the interference range is $R_i^I\left(p_i\right)=\left(p_i \cdot g_0/\beta\right)^{1/\gamma}$.

Next, we define the optimization problem to be solved every time a new application arrives in the system. Let $y_{ijk}$ be a binary variable indicating if sensor node $i$ is sensing test point $k$ of application $j$ and $x_i$ a binary variable indicating if sensor node $i$ is active in the network. Let $\tau^*$ be the time instant at which a new application arrives to the system and $A^*$ the set of applications that are running in the network at $\tau^*$, \emph{including} the new application, i.e.  $A^* = \left\{ a_j | \tau^* \in [\tau_j, \tau_j + \epsilon_j) \right\}$. The following sets of restrictions force all the applications in $A^*$ to be deployed. The problem may be unfeasible. If so, the system is left as it is to ensure that the current applications are not rejected.
Constraint (\ref{eq:const1}) forces all the applications in $A^*$ sense all their test points. Eq. (\ref{eq:const2}) ensures that if a sensor $i$ does not cover a test point $k$ of an application $j$, then it can not sense it. 
Eq. (\ref{eq:const3}) assures that $N_{ij}$ (maximum number of test points of the same application $j$ that a sensor $i$ can sense) is not exceeded. Eqs. (\ref{eq:const4})-(\ref{eq:const5}) are budget-type constraints for the available storage and processing load of the nodes.
  
\vspace{-4mm}

\begin{gather}
\sum_{i \in S_{jk}} y_{ijk} = 1  \qquad  \forall j \in A^{\ast}, \forall k \in T_j  \label{eq:const1} \\
y_{ijk} = 0  \qquad  \forall i \notin S_{jk}, \forall j \in A^{\ast}, \forall k \in T_j  \label{eq:const2} \\
\sum_{k \in T_j} y_{ijk} \leq N_{ij}  \qquad \forall i \in S,  \forall j \in A^{\ast}  \label{eq:const3} \\
\sum_{j \in A^{\ast}} \sum_{ k \in T_j} m_j y_{ijk} \leq M_i  \qquad   \forall i \in S  \label{eq:const4} \\
\sum_{j \in A^{\ast}} \sum_{ k \in T_j} l_j y_{ijk} \leq L_i  \qquad   \forall i \in S  \label{eq:const5}
\end{gather}

Deployed applications will most likely require that data generated locally are delivered remotely to collection points (sink nodes) through multihop paths. 
By resorting to a fluid model, it should be ensured that all the data produced by the sensors are received by the sink nodes. This fact can be expressed using constraints (\ref{eq:const6})-(\ref{eq:const8}). Constraint set (\ref{eq:const9}) enforces that if a sensor node is either running an application or receiving data, then it must be active in the network. Constraints (\ref{eq:const10})-(\ref{eq:const12}) ensure that all the traffic flowing out of a sensor has only one possible route to a sink, as it is typical in WSNs routes.

\vspace{-3mm}

\begin{gather}
f_{ih} = \sum_{j \in A^{\ast}} f_{ihj}  \quad    \forall i, h \in S 
\label{eq:const6} \\
\sum_{\substack{ h \in S \\ i \neq h }}f_{hij} - \sum_{\substack{h \in S \\ h \neq i} }f_{ihj} + \sum_{k \in T_j } c_j y_{ijk} = 0, \forall j \in A^{\ast}, \forall i \in S'
\label{eq:const7} \\
\sum_{j \in A^{\ast}} |T_j|  c_j = \sum_{ h \in S \setminus S' } \left(\sum_{\substack{i \in S \\ i \neq h}} f_{ih}   + \sum_{j \in A^{\ast}} \sum_{ k \in T_j} c_j y_{hjk} \right)
\label{eq:const8} \\
\sum_{\substack{h \in S \\ h \neq i}}f_{hi} + \sum_{j \in A^{\ast}}\sum_{ k \in T_j } c_j y_{ijk} \leq K x_i  \qquad   \forall i \in S  
\label{eq:const9} \\
b_{ih} \leq l_{ih}   \qquad   \forall i,h \in S  \label{eq:const10} \\
\sum_{h \epsilon S} b_{ih} \leq 1   \qquad   \forall i \in S  \label{eq:const11} \\
f_{ih } \leq K b_{ih}    \qquad   \forall i,h \in S  \label{eq:const12}
\end{gather}

\noindent where $S'$ is the set of nodes that are not sinks (a subset of $S$), $f_{ihj}$ is a variable representing the flow of data of application $j$ in bps transmitted from node $i$ to node $h$, $f_{ih}$ is a variable representing the flow of data in bps transmitted from node $i$ to node $h$ and $K$ is a constant higher than the maximum transmission rate of a node. $b_{ih}$ is a binary variable which indicates if data are transmitted from node $i$ to node $h$, and $l_{ih}$ is a constant that indicates if there is a viable link between $i$ and $h$, i.e., if the distance between both nodes is less than the maximum transmission range, $l_{ih} = 1$ and $l_{ih} = 0$ otherwise.

The available bandwidth in the network is limited and must be shared among sensor nodes. We assume that a fair medium access control scheme orchestrates the access. 
Given a directional link between a pair of nodes $\left(i,h\right)$, let the capacity of the link be defined as $C_{ih} = \min\left(C_i, C_h\right)$. 
According to the considered protocol interference model, for each link in the network it must be ensured that the fraction of time used by the link plus all its interferences is less or equal to 1:

\vspace{-3mm}

\begin{gather}
\label{eq:const13} 
\frac{f_{ih}}{C_{ih}} + \sum_{\substack{g \in S \\ g \neq h}} \frac{f_{ig}}{C_{ig}} +  \sum_{g \in S} \frac{f_{gi}}{C_{gi}} + \sum_{\substack{g \in S \\ g\neq i}} \frac{f_{hg}}{C_{hg}} +  \sum_{\substack{g \in S \\ g\neq i}} \frac{f_{gh}}{C_{gh}} + \nonumber \\   \sum_{\substack {g,t \in S \\ d_{it}<R_i^I(p_i) }} \frac{f_{gt}}{C_{gt}} + \sum_{\substack {g,t \in S \\ d_{gh}<R_g^I(p_g) }} \frac{f_{gt}}{C_{gt}} \leq 1  \qquad   \forall i,h \in S  
\end{gather}

The energy budget of each node $i$, $E_i$, is limited and it decreases when an application is deployed or the node has to forward data from other nodes. 
The power dissipation for the application $j$ at the radio transmitter $P^{t}_{ij}$ or at the radio receiver $P^{r}_{ij}$ of each node $i$  can be modeled as \cite{Hou08}: 

\vspace{-3mm}

\begin{gather}
P^{t}_{ij} = \sum_{h \in S, h \neq i} \left(\beta_1 + \beta_2  d^{\gamma}_{ih} \right)  f_{ihj}   \qquad   \forall i \in S, \forall j \in A^{\ast} 
\label{eq:powertxall}
\end{gather}
\begin{gather}
P^{r}_{ij} = \rho  \sum_{h \in S, h \neq i} f_{hij}   \qquad   \forall i \in S, \forall j \in A^{\ast} 
\label{eq:powerrxall}
\end{gather}

Typical values for $\beta_1$, $\beta_2$ and $\rho$ are $ \beta_1 =\rho = 50$ nJ/bit and  $\beta_2 = 0.0013 \text{pJ/bit/m}^4$,  with $\gamma = 4$ the path loss index. 

The estimation of the power dissipation due to the processing load, which can not be neglected in multimedia applications \cite{RedondiTMC2016}, depends on factors such as the hardware architecture or the specific application implementation. Because of this, in Eq. (\ref{eq:constf110}) it is left as a function $f$ of the processing loads $l_j$ of the applications. 
Additionally, we also consider that there is a cost $\varphi$ incurred every time a node is activated. This cost is related to the amount of energy that the node needs to wake up from the sleep mode. We also allow moving applications from one node to another as long as all the restrictions described previously are fulfilled. Nevertheless, we assume that moving an application has a cost $\delta$ due to the impact of receiving the application bytecode in the new node. With this, the energy constraints that must be ensured in every node are:

\vspace{-5mm}

\begin{gather}
\varphi \cdot (x_{i}-X_{i})\cdot x_{i} + \delta \sum_{j \in A} \sum_{ k \in T_{j}} (y_{ijk}-Y_{ijk})\cdot y_{ijk} + \nonumber \\ \sum_{j \in A} P^{t}_{ij} \Delta \tau_j +  \sum_{j \in A} P^{r}_{ij} \Delta \tau_j +  f\left(\sum_{j \in A} \sum_{ k \in T_{j}} y_{ijk} l_j\right) \Delta \tau_j + \lambda_i \nonumber \\ = E_i(\tau^*) \qquad  \lambda_i \geq 0 \qquad  \forall i \in S'
\label{eq:constf110}
\end{gather}

\noindent where $X_i$ is a constant equal to 1 if the node $i$ was active before the arrival of the new application, and 0 otherwise. $Y_{ijk}$ is a constant equal to 1 if the test point $k$ of the application $j$ was being sensed in the node $i$ just before the arrival of the new application, and 0 otherwise. $\Delta \tau_j$ is the remaining lifetime of application $j$ at time instant $\tau^*$ ($\Delta \tau_j = \tau_j + \epsilon_j - \tau^*$), $E_i(\tau^*)$ is the remaining energy that node $i$ has at $\tau^*$, and $\lambda_i$ is a variable indicating the residual energy that node $i$ would have once the lifetime of the applications deployed in it or forwarded by it expires. We assume the sinks do not have energy constraints since they can be plugged directly into the grid.

If the solution space described by those restrictions is null, then the new application cannot be deployed ensuring the presence of the previous applications and therefore the system rejects it. If the solution space contains several feasible solutions, we should select the solution that maximizes the capacity of the system of accepting future applications. To do so, we proposed three possible objective functions for the optimization problem: The first one, denoted as \textit{Total} is to maximize the total residual energy of the network (\ref{eq:maximiza1}). The second one, denoted as \textit{Max-min} is to maximize the residual energy of the node with the lowest energy (\ref{eq:maximiza2}). Finally, we also consider a weigthed sum of the two previous alternatives (\ref{eq:maximiza3}), denoted as \textit{Mixed}:

\vspace{-5mm}

\begin{gather}
\max \sum_{i \in S'} \lambda_i
\label{eq:maximiza1} \\
\max \lambda \qquad \lambda \leq \lambda_i \qquad \forall i \in S'
\label{eq:maximiza2} \\
\max \left( \lambda + \frac{1}{|S'|}\sum_{ i \in S'} \lambda_i \right)
\label{eq:maximiza3}
\end{gather}

\section{Performance evaluation}
\label{performance}

As a reference, we have focused on \emph{multimedia} applications, which require the sensing, processing and delivery of multimedia content, specifically, in visual sensor networks, i.e. WSNs designed to perform visual analysis \cite{RedondiTMC2016}. Based on the characterization of that work, the requirements vector $r_j = \left\{\text{12 kb/s, 842 KB, 69.23 MIPS}\right\}$ is used to represent the applications and the associated power dissipation (function $f$ in eq. (\ref{eq:constf110})) is 0.2 W. To support these applications, we consider \emph{high-level} sensor node hardware. The parameters have been derived by taking as a reference BeagleBone platforms \cite{beagle} equipped with an IEEE 802.15.4 radio and a low-power USB camera. The resource vector is  $o_i = \left\{\text{250 kb/s, 256 MB, 720 MIPS, 32400 J}\right\}$ assumming 2 AA batteries for all the nodes except sinks, which can be plugged directly into the grid. Results have been obtained by solving the optimization model with CPLEX software \cite{cplex}, averaging the outcome over 100 random instances of the scenario. In each instance, 200 visual applications are generated in a scenario with 36 BeagleBone nodes according to a Poisson process with rate of $1$ application per hour and a constant lifetime $\epsilon_j$ of $5$ hours. The number of test points per application is $3$, $N_{i,j} = 1$ and there are $2$ sink nodes. Nodes are deployed in a $141 \times 141$ m scenario. The sensing range is $R_i^{s} = 40$ m. A path loss model with $\gamma = 4$ and $g_0 = 8.1 \cdot 10^{-3}$ is considered. $P_{max}= -10$ dBm, $\alpha = -92$ dBm and $\beta = -104$ dBm.

\begin{figure*}[!t]
	\centering
	{
		\subfigure[]{\includegraphics[width=2.28in]{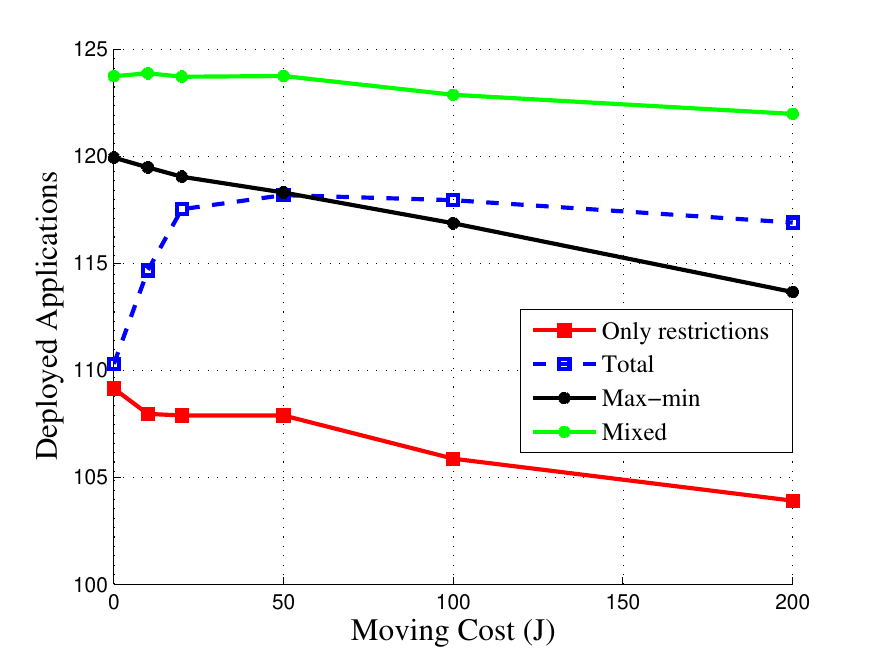}
			\label{fig:ACONApps}}
		\subfigure[]{\includegraphics[width=2.28in]{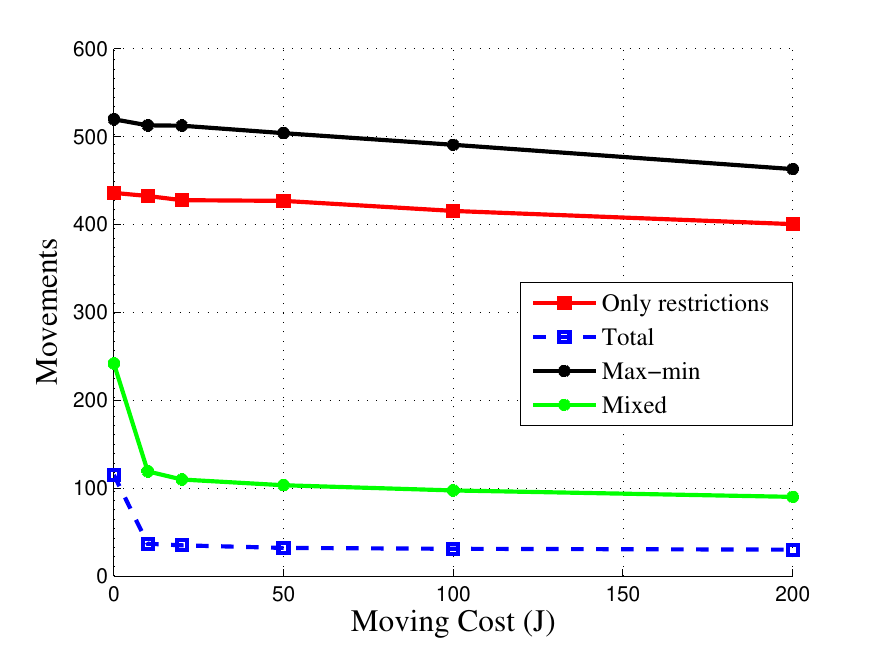}
			\label{fig:ACMov}}
		\subfigure[]{\includegraphics[width=2.28in]{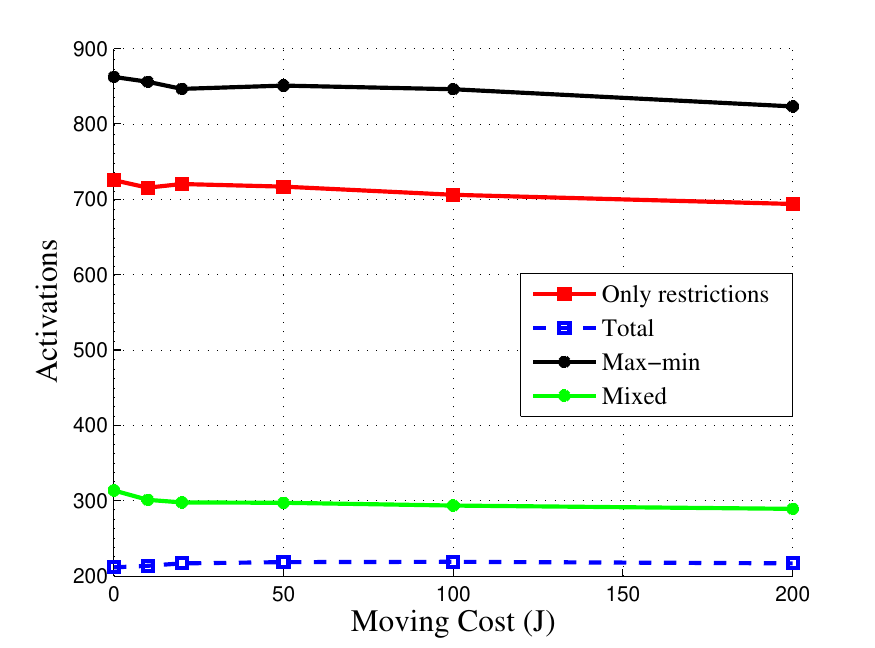}
			\label{fig:ACActiv}}

		\vspace{-3mm}
		\caption{Impact of moving cost $\delta$. a) Deployed applications. b) Number of movements c) Number of activations. $\varphi$ = 10~J.}
		\label{fig:AC}
	}
	\end{figure*}
	
	\begin{figure*}[!t]
	
	\vspace{-5mm}
	
		\centering
		{
			\subfigure[]{\includegraphics[width=2.28in]{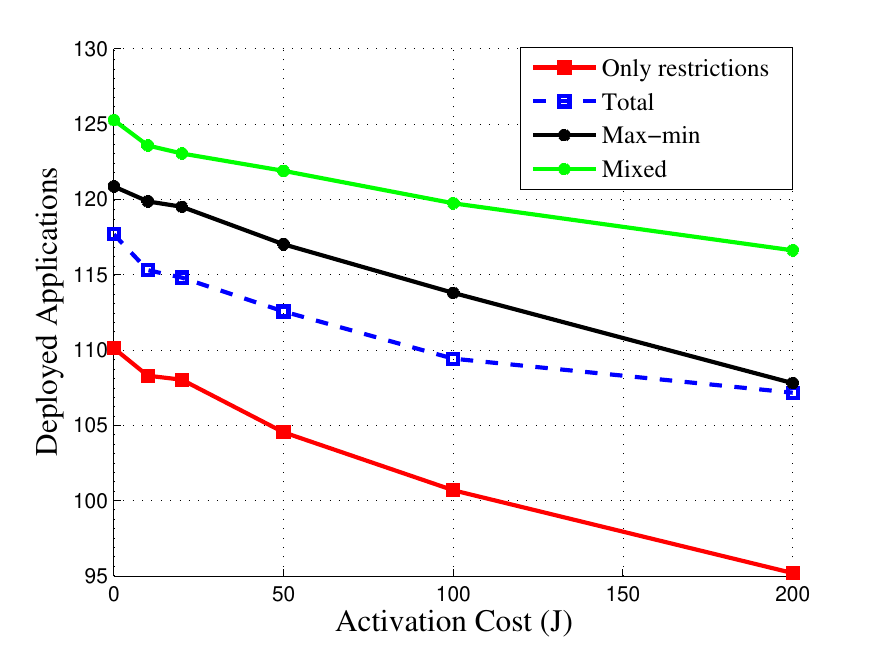}
				\label{fig:SCONApps}}
			\subfigure[]{\includegraphics[width=2.28in]{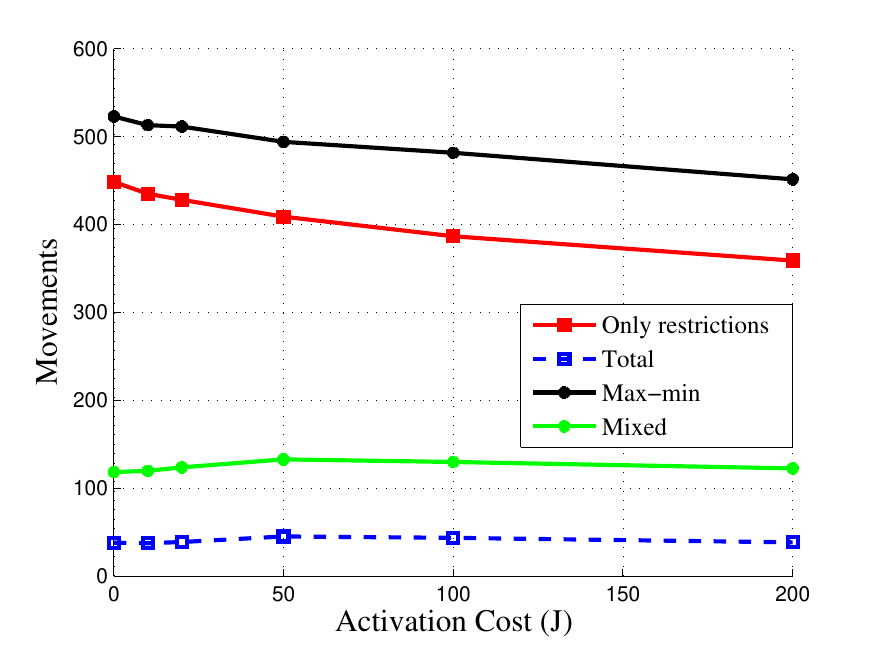}
				\label{fig:SCMov}}
			\subfigure[]{\includegraphics[width=2.28in]{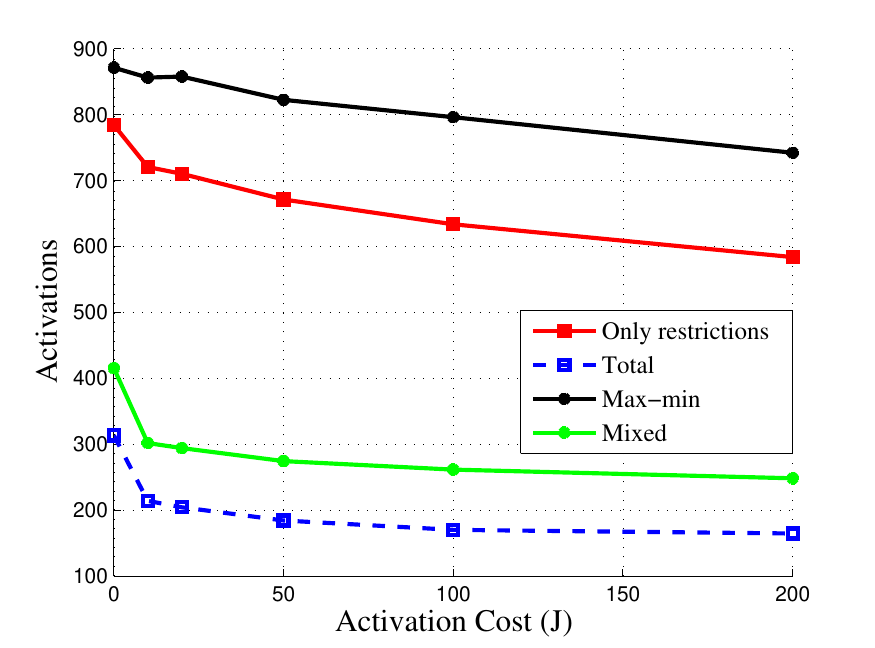}
				\label{fig:SCActiv}}
			\vspace{-3mm}
			\caption{Impact of activation cost $\varphi$. a) Deployed applications. b) Number of movements c) Number of activations. $\delta$ = 10~J.}

			\vspace{-6mm}			
			
			\label{fig:SC}
			
		}
	\end{figure*}

In addition to the three proposed objective functions, results are also presented for the case where no objective function is considered (\textit{Only restrictions}). 
Fig.~\ref{fig:ACONApps} shows that the total number of deployed applications is always higher for the \textit{Mixed} approach; as expected the worst results are obtained with the \textit{Only restrictions} approach and finally depending on the value of the moving cost, \textit{Total} and \textit{Max-min} strategies outperform each other. For the \textit{Total} strategy, initially, as the moving cost increases, the number of deployed applications rises. This can be explained as follows: when the moving cost is low and a new application arrives at the system, the model prefers to move one current application from one active node to another, rather than activating a new node. This makes new applications tend to be located in the already activated nodes, leading to these nodes running out of energy faster. In the end, this makes the network disjoint and reduces the number of deployed applications. Figs.~\ref{fig:ACMov}~and~\ref{fig:ACActiv} show that \textit{Only restrictions} and \textit{Max-Min} are the strategies with more movements and activations. This is straightforward for \textit{Only restrictions}, since applications are deployed without any additional objective rather than fulfilling the constraints, and therefore the nodes where the applications are deployed are more randomly chosen. A similar explanation can be applied to the \textit{Max-Min} strategy: eq. (\ref{eq:maximiza2}) only takes care of the node with the lowest energy, so the remaining nodes can be activated or receive an application without any penalty in the objective function.
Finally, it must be noted (Fig. \ref{fig:ACActiv}) that the number of activations remains almost constant (and not rises) when the movement cost increases. This is because an activation of a new node also implies that this node has to receive the bytecode of the application, so the higher moving cost cannot be compensated by activating more nodes. 
Fig.~\ref{fig:SCONApps} shows that the \textit{Mixed} strategy keeps providing the best performance in terms of deployed applications for different values of the activation cost. Again, for the same reasons explained above, \textit{Only restrictions} and \textit{Max-Min} are the strategies that have more movements and activations (Figs~\ref{fig:SCMov} and \ref{fig:SCActiv}). In addition, it is worth noting that for \textit{Mixed} and \textit{Total} approaches, the number of movements increases as the activation cost rises to minimize the energy consumption in the nodes.

\section{Conclusion}
\label{conclusion}

In this paper we have analyzed how to dynamically allocate the resources of a shared sensor network to multiple applications. Namely, the proposed optimization framework accounts for constraints on the sensor nodes capabilities  and network limitations, including additional energy consumption related to resource re-allocation. Different alternatives related to the residual node energy have been analyzed as objective function: total residual energy, max-min of node energy and a mix of both metrics. The results obtained for realistic network scenarios show that the mixed metric provides the best performance in terms of number of deployable applications.

\section*{Acknowledgment}

This work has been supported by the Spanish Government through the grant TEC2014-52969-R from the Ministerio de Ciencia e Innovaci\'on (MICINN), Gobierno de Arag\'on (research group T98), the European Social Fund (ESF), Centro Universitario de la Defensa through project CUD2013-05, Universidad de Zaragoza, Fundaci\'on Bancaria Ibercaja and Fundaci\'on CAI (IT 2/15). This work has also been partially supported by the Italian Ministry for Education, University and Research (MIUR) through the national cluster project SHELL, Smart Living technologies (grant number: CTN01 00128 111357).

\footnotesize
\bibliographystyle{IEEEtran}
\bibliography{./IEEEabrv,./bare_conf}

\begin{thebibliography}{10}
\providecommand{\url}[1]{#1}
\csname url@samestyle\endcsname
\providecommand{\newblock}{\relax}
\providecommand{\bibinfo}[2]{#2}
\providecommand{\BIBentrySTDinterwordspacing}{\spaceskip=0pt\relax}
\providecommand{\BIBentryALTinterwordstretchfactor}{4}
\providecommand{\BIBentryALTinterwordspacing}{\spaceskip=\fontdimen2\font plus
\BIBentryALTinterwordstretchfactor\fontdimen3\font minus \fontdimen4\font\relax}
\providecommand{\BIBforeignlanguage}[2]{{%
\expandafter\ifx\csname l@#1\endcsname\relax
\typeout{** WARNING: IEEEtran.bst: No hyphenation pattern has been}%
\typeout{** loaded for the language `#1'. Using the pattern for}%
\typeout{** the default language instead.}%
\else
\language=\csname l@#1\endcsname
\fi
#2}}
\providecommand{\BIBdecl}{\relax}
\BIBdecl

\bibitem{Khan15}
I.~Khan, F.~Belqasmi, R.~Glitho, N.~Crespi, M.~Morrow, and P.~Polakos, ``Wireless sensor network virtualization: Early architecture and research perspectives,'' \emph{IEEE Network}, vol.~29, pp. 104 -- 112, may-jun. 2015.

\bibitem{Madria14}
S.~Madria, V.~Kumar, and R.~Dalvi, ``Sensor cloud: A cloud of virtual sensors,'' \emph{IEEE Software}, vol.~31, no.~2, pp. 70 -- 77, mar.-apr. 2014.

\bibitem{Xu2010}
Y.~Xu, A.~Saifullah, Y.~Chen, C.~Lu, and S.~Bhattacharya, ``Near optimal multi-application allocation in shared sensor networks,'' in \emph{Proceedings of the Eleventh ACM International Symposium on Mobile Ad Hoc Networking and Computing MobiHoc}.\hskip 1em plus 0.5em minus 0.4em\relax ACM, 2010, pp. 181--190.

\bibitem{Zeng2015TC}
D.~Zeng, P.~Li, S.~Guo, T.~Miyazaki, J.~Hu, and Y.~Xiang, ``Energy minimization in multi-task software-defined sensor networks,'' \emph{Computers, IEEE Transactions on}, vol.~64, no.~11, pp. 3128--3139, Nov 2015.

\bibitem{Delgado2015}
C.~Delgado, J.~R. G{\'{a}}llego, M.~Canales, J.~Ort{\'{\i}}n, S.~Bousnina, and M.~Cesana, ``An optimization framework for resource allocation in virtual sensor networks,'' in \emph{{IEEE} {GLOBECOM} 2015, San Diego, CA, USA, December 6-10, 2015}, pp. 1--7.

\bibitem{Delgado2016}
C.~Delgado, J.~R. G{\'a}llego, M.~Canales, J.~Ort{\'\i}n, S.~Bousnina, and M.~Cesana, ``On optimal resource allocation in virtual sensor networks,'' \emph{Ad Hoc Networks}, vol.~50, pp. 23--40, 2016.

\bibitem{Hou08}
T.~Hou, Y.~Shi, and H.~Sherali, ``Rate allocation and network lifetime problems for wireless sensor networks,'' \emph{IEEE/ACM Transactions on Networking}, vol.~16, no.~2, pp. 321 -- 334, 2008.

\bibitem{RedondiTMC2016}
A.~Redondi, L.~Baroffio, L.~Bianchi, M.~Cesana, and M.~Tagliasacchi, ``Compress-then-analyze vs analyze-then-compress: what is best in visual sensor networks?'' \emph{IEEE Transactions on Mobile Computing}, vol.~PP, no.~99, pp. 1--1, 2016.

\bibitem{beagle}
G.~Coley, \emph{Beaglebone rev a6 system reference manual}, 2012.

\bibitem{cplex}
``{ILOG CPLEX},'' \url{http://www-01.ibm.com/software/integration/optimization/cplex-optimizer/}, 2015.

\end{thebibliography}

\end{document}